%% file: main.tex
\documentclass[lettersize,journal]{IEEEtran}
\input{macro}

\begin{document}

\title{SAM Audio Judge: A Unified Multimodal Framework for Perceptual Evaluation of Audio Separation}

\author{Helin Wang$^{*}$, Bowen Shi$^{\dagger}$, Andros Tjandra, John Hoffman, Yi-Chiao Wu, Apoorv Vyas, \\Najim Dehak, Ann Lee and Wei-Ning Hsu$^{\dagger}$
\thanks{$^{*}$Work done at Meta. (email: hwang258@jhu.edu)}
\thanks{
Helin Wang and Najim Dehak are with the Johns
Hopkins University.}
\thanks{
Bowen Shi, Andros Tjandra, John Hoffman, Yi-Chiao Wu, Apoorv Vyas, Ann Lee and Wei-Ning Hsu are with Meta.
}
\thanks{$^{\dagger}$Wei-Ning Hsu and Bowen Shi are coressponding authors. (email: wnhsu@meta.com, bshi@meta.com)
}
}

\maketitle

\begin{abstract}
The performance evaluation remains a complex challenge in audio separation,
and existing evaluation metrics are often misaligned with human perception, course-grained, relying on ground truth signals. On the other hand, subjective listening tests remain the gold standard for real-world evaluation, but they are expensive, time-consuming, and difficult to scale. 
This paper addresses the growing need for automated systems capable of evaluating audio separation without human intervention. 
The proposed evaluation metric, SAM Audio Judge (SAJ), is a multimodal fine-grained reference-free objective metric, which shows highly alignment with human perceptions. SAJ supports three audio domains (speech, music and general sound events) and three prompt inputs (text, visual and span), covering four different dimensions of evaluation (recall, percision, faithfulness, and overall).
SAM Audio Judge also shows potential applications in data
filtering, pseudo-labeling large datasets and reranking in audio separation models.
We release our code and pre-trained models at: \url{https://github.com/facebookresearch/sam-audio}.
\end{abstract}


\input{introduction}

\input{related_works}

\input{text_saj}

\input{multimodal_saj}

\input{conclusion}

\input{ack}

\bibliographystyle{IEEEtran}

\bibliography{ref}

\vfill

\end{document}

%% file: macro.tex
\usepackage{amsmath,amsfonts}
\usepackage{algorithmic}
\usepackage{algorithm}
\usepackage{array}
\usepackage{textcomp}
\usepackage{stfloats}
\usepackage{url}
\usepackage{verbatim}
\usepackage{graphicx}
\usepackage{cite}

\usepackage[utf8]{inputenc} 
\usepackage[T1]{fontenc}    
\usepackage{hyperref}       
\usepackage{url}            
\usepackage{booktabs}       
\usepackage{amsfonts}       
\usepackage{nicefrac}       
\usepackage{float}
\usepackage{etoolbox}
\usepackage{xspace}
\usepackage{adjustbox}
\usepackage[table,dvipsnames]{xcolor}  
\usepackage{amssymb}
\usepackage{tabularx}
\usepackage{longtable}
\usepackage{booktabs}
\usepackage{threeparttable}
\usepackage{array}
\usepackage{pifont}
\usepackage{multirow}
\usepackage{wrapfig}

\usepackage{microtype}      
\usepackage{xcolor}         
\usepackage{multirow}
\usepackage{graphicx}
\usepackage{subcaption}
\usepackage{pifont}    
\usepackage{tcolorbox}

\usepackage{musicography}
\usepackage{tikz}
\usepackage{multicol}
\usepackage{enumitem}

\definecolor{MyLightBlue}{HTML}{0072BD}
\definecolor{MyLightPurple}{HTML}{7E2F8E}
\definecolor{MyLightRed}{HTML}{D95319}
\definecolor{MyLightGreen}{HTML}{77AC30}

\definecolor{SystemBackBase}{HTML}{F5F5F5}
\definecolor{SystemFrame}{HTML}{666666}
\definecolor{UserBackBase}{HTML}{DAE8FC}
\definecolor{UserFrame}{HTML}{6C8EBF}
\definecolor{LLMBackBase}{HTML}{E1D5E7}
\definecolor{LLMFrame}{HTML}{9673A6}

\colorlet{SystemBack}{SystemBackBase!10}
\colorlet{UserBack}{UserBackBase!10}
\colorlet{LLMBack}{LLMBackBase!10}

%% file: introduction.tex
\section{Introduction}

Traditional evaluation of audio separation systems relies on distortion-based metrics such as signal-to-distortion ratio (SDR), scale-invariant SDR (SI-SDR), signal-to-interference ratio (SIR) and signal-to-artifact ratio (SAR)~\cite{vincent2006performance,le2019sdr}. These measures quantify energy differences between separated outputs and reference signals, and have been widely adopted in benchmark datasets such as WSJ0-2mix~\cite{hershey2016deep} and MUSDB18~\cite{stoter2019open}. However, these metrics require access to the ground-truth clean signal, which is rarely available in real-world mixtures. Moreover, they provide limited perceptual insight: for example, two outputs with nearly identical SDR values may sound drastically different~\cite{le2019sdr}, and such distortion-based metrics are known to correlate poorly with human mean opinion scores (MOS)~\cite{cartwright2018crowdsourced,cano2016evaluation}.

Subjective listening tests remain the gold standard for evaluation~\cite{series2014method}, but they are expensive, time-consuming, and difficult to scale. This creates a persistent gap between easily computed distortion errors and perceptually meaningful assessments. To narrow this gap, perceptual metrics originally designed for speech coding and transmission, such as POLQA~\cite{beerends2013perceptual} and ViSQOL~\cite{hines2015visqol}, attempt to model auditory mechanisms. While effective in their intended domains, they generalize poorly to the diverse artifacts produced by modern audio systems~\cite{delgado2024towards}.

More recently, data-driven quality predictors have gained traction, which have shown stronger alignment with human judgments~\cite{huang2022voicemos,chinen2020visqol,mittag2021nisqa,manocha2021cdpam,reddy2021dnsmos,tjandra2025meta}. Nonetheless, most of these efforts focus on speech synthesis, enhancement, or audio generation. In audio separation, evaluation still relies mainly on SDR-like metrics, and listening studies~\cite{jaffe2025musical} showed their poor correspondence with perceptual judgments. This gap underscores the need for evaluation frameworks that move beyond distortion errors and more faithfully reflect human listening experience.
In addition, as the demand for better controllability in audio separation models grows, multiple input prompts, such as text and visual cues, are increasingly being explored. However, no existing automatic evaluation tool currently supports multi-prompt inputs.

Building on these observations, we conduct a systematic investigation into perceptually aligned evaluation for audio separation. Our goal is to design an evaluation framework that: (i) enables fine-grained perceptual assessment of separation outputs, (ii) exhibits strong correlation with human listening judgments, and (iii) supports multiple input prompts.

In this paper, we introduce \textit{i.e.} Segment Anything Model (SAM) Audio Judge, hereafter referred to as SAJ, a novel automatic reference-free evaluation tool that measures four key dimensions of multimodal-prompted audio separation performance: recall, precision, faithfulness, and overall quality. 
Compared to the text-only judge model in~\cite{samaudio}, SAJ is a full-featured separation judge model that supports arbitrary combinations of text, visual, and span prompts.
We first construct a high-quality text-prompted separation judgment dataset with human subjective listening scores. The dataset spans a wide range of audio source domains and model outputs. Using this dataset, we train text-prompted SAJ models that show strong alignment with human perceptual evaluations.
We further explore practical applications of SAJ, demonstrating its usefulness for data filtering, pseudo-labeling large-scale datasets, and reranking separation system outputs. In addition, we transfer knowledge from the text-prompted SAJ to a multimodal SAJ, resulting in a powerful evaluation model capable of handling multiple input prompts including text, visual, and span cues.

%% file: related_works.tex
\section{Related Works}

\subsection{Audio Separation}

Audio separation aims to decompose a complex sound mixture into individual source tracks associated with distinct sound events, speakers, or instruments.
Research in this area is extensive, and existing methods can be broadly categorized into promptless and prompted approaches.

Promptless systems decompose audio mixtures into a fixed set of predefined sources and have demonstrated strong performance in tasks such as speech enhancement, speaker separation, and music demixing~\cite{mitsufuji2021music,fabbro2024sound,zhao2024mossformer2,kong2023universal}.
However, these methods assume a fixed output configuration and depend on predefined taxonomies of sound categories; therefore, they struggle in open-domain scenarios or when users specify sound types whose boundaries are ambiguous or context dependent.

Motivated by these limitations, recent work has moved toward prompted separation, where the target source is specified through an external cue.
Text prompts~\cite{liu2023separate,yuan2024flowsep,Ma2024CLAPSepLC,wang2025soloaudio,jiarui2024dpmtse} allow users to describe arbitrary sound events (e.g., “dog barking,” “female speech”), removing the need for fixed category taxonomies.
However, text alone often lacks the granularity to disambiguate fine-grained or overlapping sound events. To address this, prior studies have explored audio-based conditioning~\cite{wen2025promptsep,xu2020spexplus,wang2025solospeechenhancingintelligibilityquality} as an alternative modality, while others introduced span prompting~\cite{samaudio,wang2022improvingtargetsoundextraction}, which provides temporal cues without requiring external reference audio.
Visual prompts~\cite{Zhao2018soundofpixels,Huang2024high,li2024iianet,ephrat2018looking,dong2023clipsep} further complement text by offering instance-level grounding, enabling the model to distinguish between multiple similar sources within the same scene.

\subsection{Objective Metrics}

Objective evaluation metrics play a central role in assessing the quality of audio separation systems. Traditional metrics primarily measure signal fidelity with respect to a clean ground-truth source and have been widely adopted in speech enhancement, music demixing, and universal sound separation.

The most commonly used metrics come from the SDR family, including SDR, SI-SDR, and SAR/SIR from the BSS Eval toolkit~\cite{vincent2006performance}.
These metrics decompose the estimated signal into components associated with target distortion, interference, and artifacts. They quantify how closely the separated signal matches the ground truth in a least-squares sense.
SI-SDR has become a standard benchmark due to its robustness to scale differences between estimated and target signals \cite{luo2019conv,luo2020dualpath}.
Despite their popularity, these metrics suffer from several shortcomings: (i) They require access to clean ground-truth references, which are often unavailable in real-world mixtures. (ii) They focus on sample-level reconstruction accuracy rather than perceptual quality. (iii) Empirical evidence shows that signals with similar SDR scores can sound drastically different to human listeners~\cite{le2019sdr}. (4) Correlation with human mean opinion scores (MOS) is weak, especially when distortions are perceptually subtle~\cite{cartwright2018crowdsourced,cano2016evaluation}.
These limitations make SDR-based metrics insufficient for evaluating modern open-domain, prompt-conditioned separation systems.

To mitigate the mismatch between signal fidelity and perception, recent studies explore embedding-based evaluation metrics.
For example, Fréchet Audio Distance (FAD)~\cite{kilgour2019frechet} measures the distributional distance between separated audio and reference clean audio in a learned embedding space.
Other works leverage audio classifiers (e.g., VGGish~\cite{hershey2017cnn}, CLAP~\cite{wu2023large}, WavLM~\cite{chen2022wavlm}) to compute similarity in semantic or perceptual embedding spaces.
These embedding-based metrics capture high-level perceptual attributes rather than raw waveform differences.
They can operate without temporal alignment between estimated and target signals, and reflect semantic consistency more effectively than SDR.
However, these methods still rely on either (i) domain-matched reference datasets, or (ii) learned feature spaces not explicitly optimized for separation quality assessment.
They therefore remain imperfect surrogates for human perceptual judgments.

Another line of research evaluates separation quality using downstream task performance.
Examples include automatic speech recognition accuracy~\cite{wang2019voicefilter,barker2015third}, speaker identification accuracy~\cite{delcroix2018single}, or sound event detection performance~\cite{mesaros2017dcase,turpault2019sound} on separated signals.
Such metrics provide insight into how well separation preserves semantically meaningful information needed for downstream models. Nevertheless, task-based evaluations depend heavily on the choice of downstream system and may not reflect general perceptual quality. Their interpretability is limited, and they do not measure artifacts, distortions, or perceptual naturalness.

Overall, existing objective metrics capture different facets of separation quality, but none adequately reflect fine-grained perceptual dimensions such as timbral faithfulness, artifact severity, or semantic correctness.
In addition, nearly all existing metrics assume promptless separation and cannot accommodate multi-prompt settings (e.g., text-prompted or visually grounded separation).

\subsection{Subjective Metrics}

Subjective listening tests remain the most reliable way to evaluate audio separation because they directly capture human perceptual judgments.
The Mean Opinion Score (MOS) is the most widely used protocol, where listeners rate samples on a 1–5 scale~\cite{cartwright2016fast,cartwright2018crowdsourced,liu2023separate}.
For higher-resolution assessment, MUSHRA~\cite{schoeffler2018webmushra} presents multiple system outputs alongside references and anchors, enabling finer comparisons.
Pairwise listening tests such as AB/ABX are also commonly used to compare subtle differences between systems.

Crowdsourcing platforms (e.g., Mechanical Turk, Prolific) have made subjective evaluation more scalable, but these methods remain costly, time-consuming, and difficult to apply at large scale.
Moreover, subjective tests do not naturally support prompt-conditioned separation, where correctness must be evaluated relative to the user’s prompt.

These limitations motivate the need for automatic, perceptually aligned evaluation tools that maintain the benefits of human listening while offering scalability and prompt-awareness.

%% file: text_saj.tex
\section{Text-prompted SAJ}
\subsection{Data collection}
\label{sec:text_saj_data}

Existing evaluation guidelines for audio separation are typically simplistic, focusing only on coarse criteria such as the relevance between the separated audio and the prompt, or the overall audio quality of the output \cite{lass}. Such high-level and loosely defined objectives make the evaluation ambiguous. Scores collected under these settings often conflate multiple perceptual aspects and may be biased toward certain criteria depending on the raters’ individual interpretations, resulting in outcomes that are difficult to interpret consistently. In addition, existing studies rarely examine the difficulty of audio separation tasks themselves. Most prior work has focused solely on evaluating model outputs, without systematically analyzing how intrinsic factors such as the number of overlapping sources, loudness imbalance, or acoustic similarity between sounds affect human perception of task difficulty. Understanding separation difficulty is crucial, as it provides insights into the limitations and robustness of current models, facilitates the curriculum design for training and evaluation, and enables difficulty-aware benchmarking and adaptive model selection in real-world applications.

To better characterize the performance of audio separation models and the intrinsic difficulty of audio separation tasks, we introduce a new human annotation guideline, which defines nine perceptual dimensions.

The SAJ performance dimensions evaluate how well a model separates the target sounds:
\begin{itemize}
    \item \textbf{Recall}: Does the extracted audio contain all of the target sounds specified in the prompt?
    \item \textbf{Precision}: How effectively does the model remove non-target sounds from the extracted audio?
    \item \textbf{Faithfulness}: For target sounds present in the extracted audio, how similar do they sound to their counterparts in the original mixture?
    \item \textbf{Overall quality}: What is the overall perceptual quality of the model’s output?
\end{itemize}
In addition, the SAJ difficulty dimensions assess the complexity of the separation task itself:
\begin{itemize}
    \item \textbf{Counting}: How many non-target sounds are present in the source audio?
    \item \textbf{Overlapping}: To what extent do the target sounds overlap with non-target sounds?
    \item \textbf{Loudness}: How loud are the target sounds relative to the non-target sounds?
    \item \textbf{Confusion}: How easily can the non-target sounds be mistaken for the target sounds?
    \item \textbf{Overall difficulty}: Considering all the above factors, how difficult is it to extract the target sounds from the mixture?
\end{itemize}

\input{tables/saj_model}

Based on the above definitions, we design an annotation task to collect SAJ data, where human raters evaluate the nine axes using a five-point Likert scale (1–5). We first focus on \textbf{text-prompted} SAJ, the most commonly used scenario. We provide a comprehensive annotation guideline, which offers detailed explanations of each axis and specifies fine-grained aspects to consider during evaluation. To help raters calibrate their judgments, the guideline is supplemented with numerous audio examples and score references that illustrate what constitutes high or low scores along each dimension.

\input{tables/saj_data}

\textbf{Paired data preparation.} We collect a comprehensive set of datasets spanning music, speech, and sound effects. To mitigate the mismatch between real-world and simulated data, we use both real mixtures and synthetic mixtures as input audio. Table~\ref{tab:saj_data} shows the detailed data statistics.
For each data source, we adopt either the original sound annotations or the sound type predictions generated by the PLM-Audio model~\cite{samaudio} as text prompts. We adopt the same text prompt settings as SAM Audio~\cite{samaudio}, which support speech separation, speaker separation, music separation, instrument separation, and general sound separation. Following~\cite{jiarui2024dpmtse,wang2025soloaudio}, the text prompts are designed as single-sound descriptions, which may refer to either a specific sound source (e.g., dog barking, guitar) or a general sound category encompassing multiple sources (e.g., dog, music playing). All text is lower-cased during preprocessing.
Compared with the compositional textual descriptions proposed in~\cite{lass}, such as “a bird is chirping under a thunderstorm”, we decompose such cases into separate prompts, “bird chirping” and “thunderstorm”, which better reflect practical use scenarios.
For each modality, we gather outputs from various open-source audio separation models, which are listed in Table~\ref{tab:saj_model}.
For promptless models such as MossFormer2 and Demucs, we obtain all separated tracks and use CLAP scores~\cite{wu2023large} to associate each track with the corresponding text prompt.

We design a rater qualification program to ensure the selection of high-quality annotators. Specifically, we curate a 40-sample golden set, with scores labeled by experts and treated as ground truth. We then invite outsourced vendors to annotate this golden set and qualify their annotators using a Bayesian rater modeling approach. This model estimates a posterior distribution over each annotator’s confusion matrix, capturing their reliability and bias across rating values. From this posterior, we derive the expected Mean Absolute Deviation (MAD) of each annotator’s scores, which naturally accounts for partially completed tasks, item-score uncertainty, and imbalances in score distributions.
Annotators with the lowest expected MAD on the overall score are selected as qualified raters. While we do not directly compare against expert annotations, we use them as a spot check to ensure that the selected annotators exhibit reasonable consensus and consistent rating behavior. In the end, we successfully recruit 128 qualified raters, representing diverse subjective perspectives from the general public.

To establish a unified SAJ score that is calibrated across different audio modalities, we shuffle audio samples from all modalities during annotation. We also apply loudness normalization to eliminate potential confounding effects introduced by variations in audio volume. Finally, we collect three independent ratings per audio sample to reduce variance and improve reliability.

\begin{figure*}[t]
  \centering
  \includegraphics[width=0.9\linewidth]{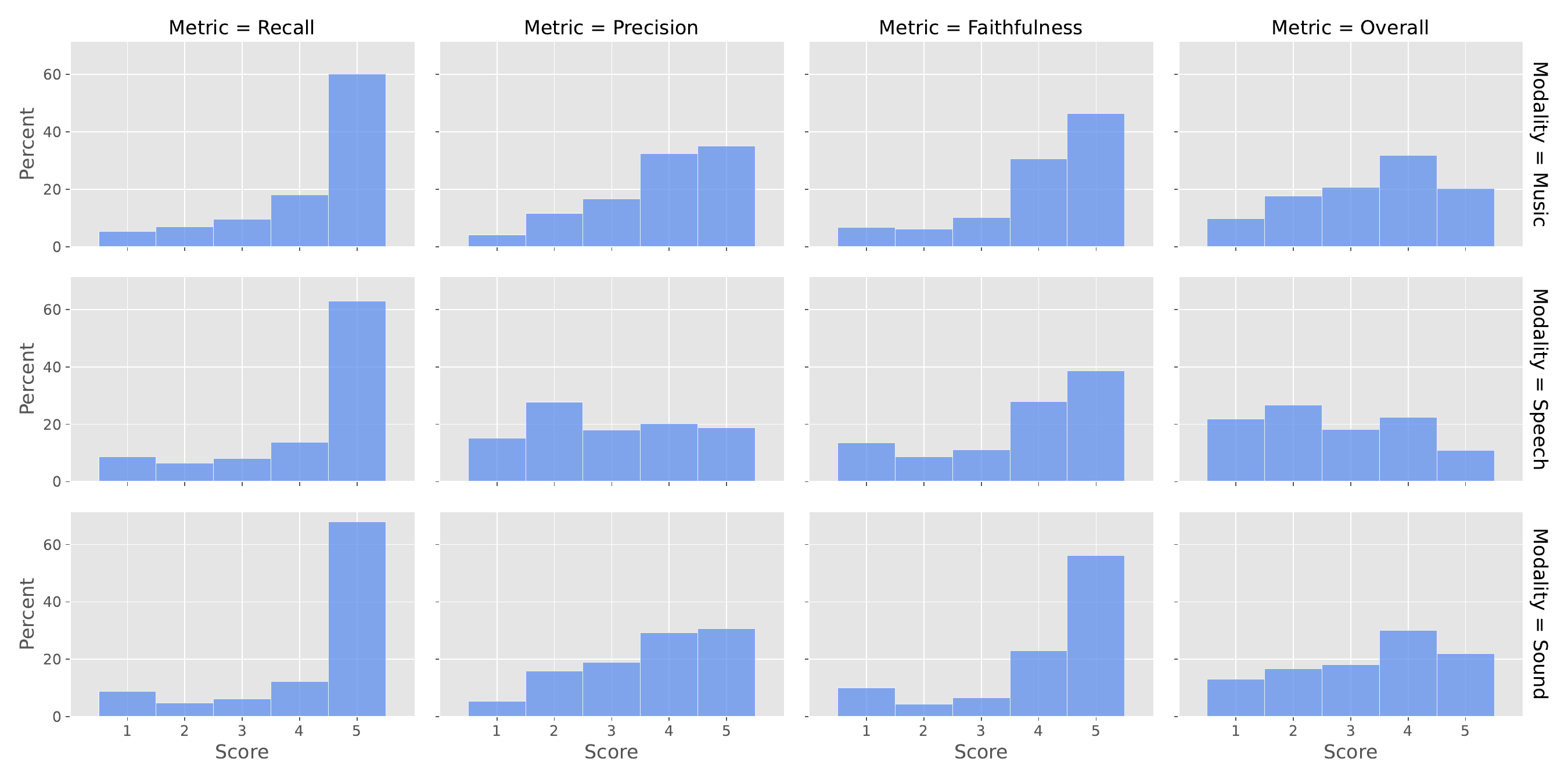}
  \caption{Distribution of recall, precision, faithfulness, and overall scores across the speech, music, and sound modalities.}
  \label{fig:saj_dataset}
\end{figure*}

\begin{figure*}[t]
  \centering
  \includegraphics[width=0.9\linewidth]{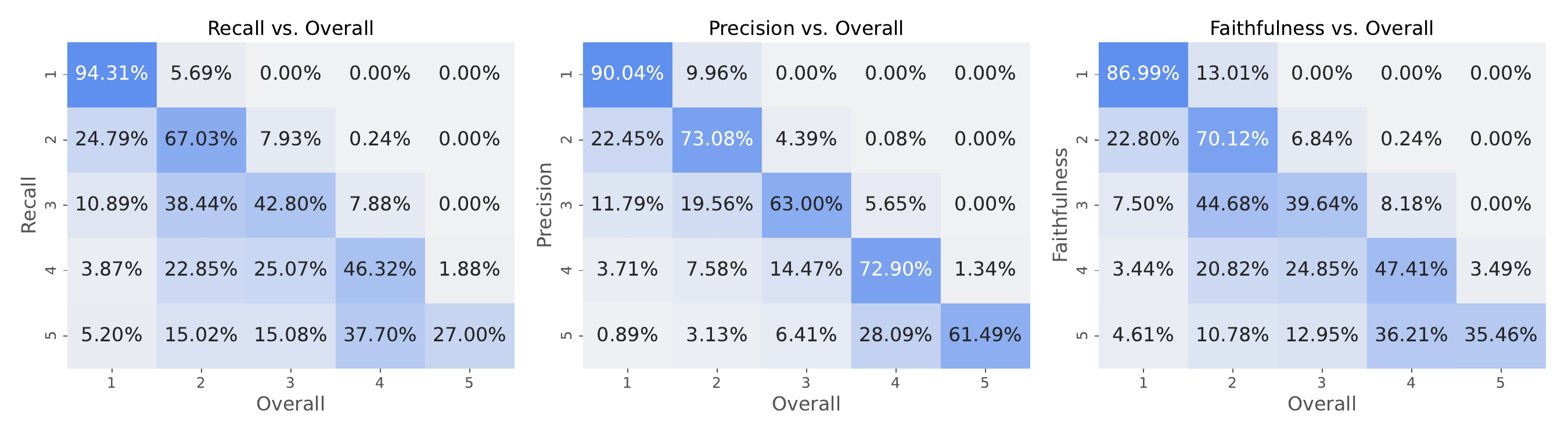}
  \caption{Joint score distributions between recall, precision, faithfulness, and overall ratings. The horizontal axes are normalized for each metric, allowing direct comparison of their correlation patterns with the overall rating.}
  \label{fig:saj_dataset_correlation}
\end{figure*}

To mitigate potential biases in training the SAJ models, we designed a three-stage data collection pipeline aimed at balancing the distribution of perceptual scores as much as possible:

\begin{itemize}
\item \textbf{Batch 1 (50 hours):}
Since no reliable objective metric exists to directly measure separation quality, we approximate performance using a combination of CLAP scores~\cite{wu2023large}, residual CLAP (the similarity difference with text between mixture and separated audio), Audiobox Aesthetics Production Complexity (AES-PC)~\cite{tjandra2025meta}, and residual AES-PC.
We compute a composite score over all samples and partition the data into three categories—\textit{poor}, \textit{medium}, and \textit{good} separation. A balanced subset is then randomly sampled from these partitions.

\item \textbf{Batch 2 (200 hours):}  
A preliminary SAJ model trained on Batch 1 is used to estimate perceptual scores for new data. We observed severe imbalance in the distributions of recall and faithfulness, with most samples concentrated at scores 1 and 5. To enrich the distribution, we augment the separation outputs through temporal masking, frequency masking, and equalization. A more balanced subset is then sampled across all four performance dimensions.

\item \textbf{Batch 3 (160 hours):}  
We analyze the distributional gaps remaining after Batches 1 and 2 and collect additional data targeted at underrepresented score ranges. A SAJ model trained on the first two batches is used to guide the sampling toward a more uniform distribution of the overall scores.

\end{itemize}

We further refined the dataset using human ratings, retaining only samples where the three annotators' scores differed by no more than 1. The remaining scores were then averaged to form the final SAJ dataset, which consists of 357.9 hours of annotated audio.
Fig.~\ref{fig:saj_dataset} presents the distribution of the four performance dimensions across speech, music, and sound. The overall scores are relatively well balanced.
Fig.~\ref{fig:saj_dataset_correlation} illustrates the pairwise correlations between recall, precision, and faithfulness with respect to the overall score. All three dimensions exhibit strong positive correlations with the overall metric.

\subsection{Model training}

\begin{figure*}[t]
  \centering
  \includegraphics[width=\linewidth]{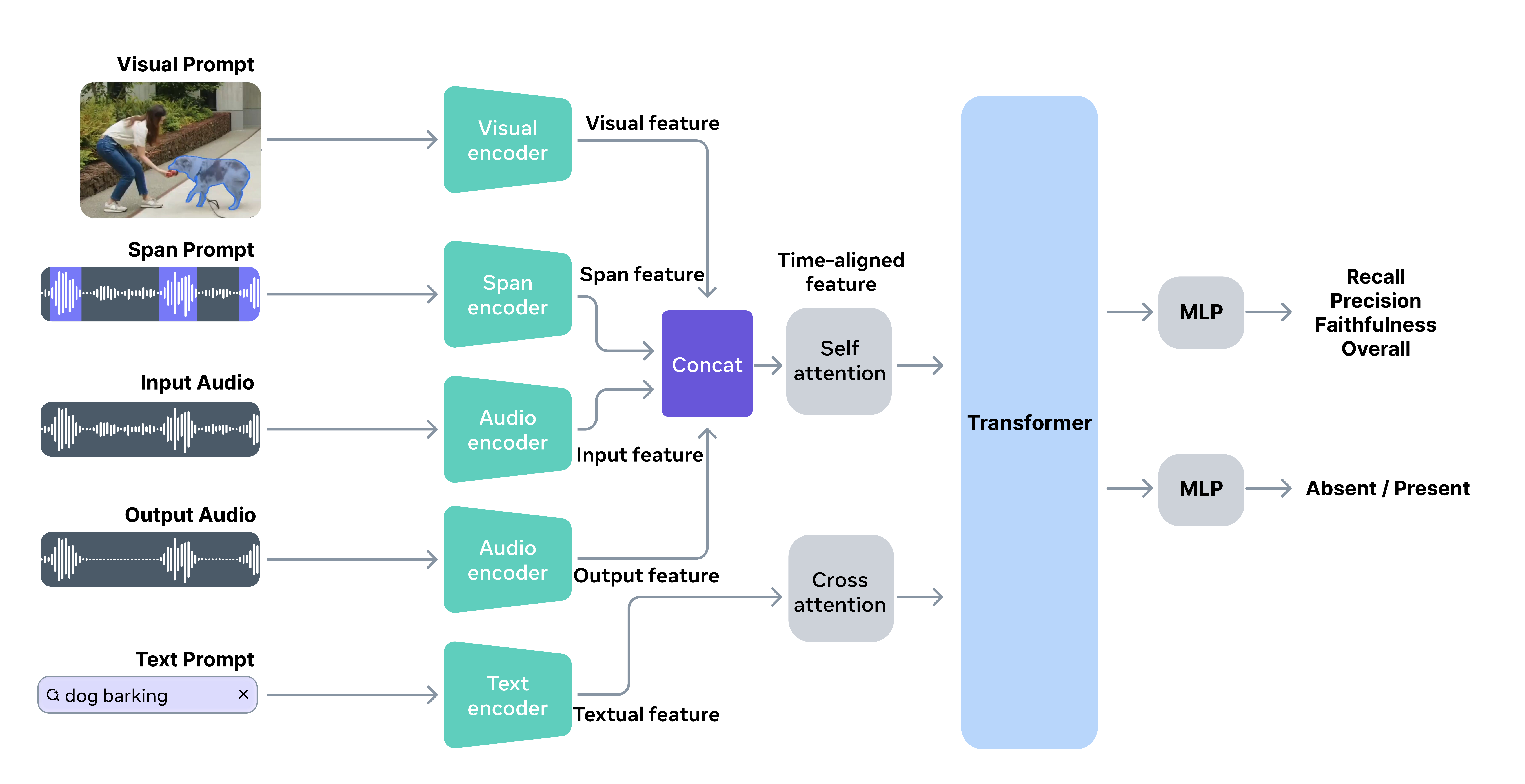}
  \caption{Overview of the SAJ model. Given the input audio and output audio, SAJ predicts the separation performance in four dimensions (recall, precision, faithfulness and overall),
conditioned on any combination of text descriptions (text prompts), visual masks (visual prompts), and temporal
intervals (span prompts).}
  \label{fig:saj_model}
\end{figure*}

The SAJ model is designed to predict the perceptual performance of separation systems as well as the difficulty of separation tasks.
As illustrated in Fig.~\ref{fig:saj_model}, SAJ takes five possible inputs: the mixture audio, the separated audio, and optionally a text prompt, a visual prompt, or a span prompt.
We employ a pretrained audio encoder to extract audio representations and a pretrained text encoder for textual features. Both encoders are adopted from PE-AV~\cite{vyas2025peav}, which is trained using large-scale audio–visual–text contrastive learning.
For visual prompting, we use the pretrained PE core~\cite{bolya2025PerceptionEncoder} as the visual encoder, and for span prompting, we use a learnable embedding layer to encode temporal spans.

We first fuse the temporal modalities: input audio features, output audio features, span feature, and visual features, using self-attention. All temporal features are resampled to match the sequence length of input audio features and then concatenated.
Next, textual features are incorporated through an additional cross-attention layer.
The resulting aligned representation is passed through a Transformer to learn joint multimodal features, followed by several linear layers that predict the SAJ scores: \textit{e.g.}, recall, precision, faithfulness, and overall.

In addition, we observe that introducing a auxiliary task, which predicts whether the separated audio follows the given prompts~\cite{wang2022detect,wang2022improving}, significantly improves model performance.
To incorporate this signal, we pretrain the entire model on a text–audio (and optionally visual–audio or span–audio) alignment detection task using a large-scale simulated dataset where ground-truth separated tracks are available.
During pretraining, the output audio is alternated between the correct target source and a randomly selected non-target source from the same mixture, enabling binary supervision for whether the target sound is present. A dedicated linear layer predicts this binary alignment.
After this pretraining stage, we fine-tune the SAJ model using human annotations to predict the SAJ scores.

We train the text-prompted SAJ models on the dataset described in Sec.~\ref{sec:text_saj_data}.
To stabilize learning, the dataset is balanced according to the overall score distribution, and the proportions of speech, music, and sound are set to 1:2:4 based on empirical performance.
We use an effective batch size of 8k audio tokens.
Both the pretraining and fine-tuning stages run for 200K updates with a constant learning rate of $3\times10^{-4}$, following a 5K-step linear warmup.
We adopt AdamW with a weight decay of 0.1 and train using \texttt{bf16} precision.
All experiments are conducted on 8 NVIDIA A100-80GB GPUs.

\subsection{Comparison with other metrics}

\input{tables/saj_com}
We compare the proposed SAM Audio Judge model with several representative baseline systems covering diverse evaluation paradigms:
\begin{itemize}
    \item CLAP~\cite{wu2023large}: A large-scale contrastive audio–text model trained on millions of audio–caption pairs. We use its cosine similarity between the text prompt and output audio embeddings as a proxy for perceptual alignment. This represents the current standard for audio–language correspondence evaluation.
    \item AES-PC diff~\cite{tjandra2025meta}: A model trained to predict subjective aesthetic preference scores for audio samples. It focuses on the complexity of audio scene, measured by the number of audio components. We calculate the AES-PC difference between input and output audios.
    \item SDR Estimator~\cite{dang2023using,frummer2025refess}:
    This baseline represents distortion-based evaluation methods that prioritize signal fidelity over perceptual quality. It is a regression model trained to approximate SDR without requiring access to ground-truth references.
    To ensure a fair comparison, we adopt the same architecture as SAJ but supervise the model using SDR values. The training set is balanced across target signal levels from $-25$ dB to $25$ dB and spans 496 hours of speech, music, and sound effects.
    The resulting SDR estimator attains PCC scores of 0.923, 0.681, and 0.665 for speech, music, and sound effects, respectively.

    \item Gemini-2.5-pro~\cite{comanici2025gemini}: A large multimodal LLM capable of reasoning over both text and audio inputs. We prompt it to rate the separated audio quality according to the same evaluation axes used in SAJ, representing a language-model-based perceptual evaluation baseline. We also provide several examples with different scores to enable few-shot learning. 
\end{itemize}
Together, these baselines cover the major paradigms in contrastive representation learning, aesthetic preference modeling, signal distortion estimation, and general-purpose multimodal reasoning, enabling a comprehensive comparison against our task-specific SAJ model.

Table~\ref{tab:res_saj_com} shows that the proposed text-prompted SAJ model consistently outperforms all baselines across speech, music, and sound under both Pearson (PCC) and Spearman (SRCC) correlation metrics.
SAJ attains markedly higher alignment with human judgments, achieving PCCs of 0.883, 0.815, and 0.815 and SRCCs of 0.817, 0.714, and 0.781 for the three modalities, respectively, which indicates strong linear agreement as well as rank-order consistency.
In comparison, baseline systems such as CLAP and Gemini-2.5-pro exhibit only moderate correlations, while SDR Estimator and AES-PC diff fail to reflect perceptual quality and often produce low correlations.
The robustness of SAJ across all modalities demonstrates its ability to generalize beyond speech to more complex acoustic domains including music and environmental sounds.
Overall, these results confirm that SAJ effectively captures human perceptual judgments through joint audio–text representations and prompt-conditioned pretraining, providing a more reliable and fine-grained evaluation framework than existing baselines.

\subsection{Ablation studies}

\input{tables/ab_loss}

\subsubsection{Training Loss} 
Table~\ref{tab:ab_loss} summarizes the impact of different training objectives on overall PCC across speech, music, and sound.
The CE and KL-divergence losses discretize the 1–5 score range into 21 bins and treat score prediction as a classification or soft-distribution matching task. However, this discretization removes the inherent continuity of the perceptual scale, leading to significantly lower correlations with human ratings across all modalities.
In contrast, regression-based losses better preserve the ordinal and continuous nature of the scores. MAE and MSE both yield substantial improvements, with MSE performing slightly better on sound effects and MAE showing more robustness on speech and music.
The best performance is achieved with a combined MAE+MSE objective, which leverages the stability of MAE against outliers and the smooth optimization behavior of MSE. This hybrid loss consistently outperforms all other objectives, indicating that preserving the continuous structure of perceptual scores is crucial for accurate separation quality prediction.

\input{tables/ab_backbone}

\subsubsection{Backbone} 
Table~\ref{tab:ab_backbone} compares the effect of different backbone encoders on overall PCC.
WavLM is a speech-only pretrained model, which provides limited perceptual alignment for separation evaluation, yielding modest correlations, particularly on music and sound effects where acoustic structures differ substantially from speech.
DAC-VAE, which compresses audio into a learned latent space, offers richer generative representations and improves performance across all modalities, but still lacks semantic grounding needed for prompt-conditioned perceptual judgments.
In contrast, PE-AV achieves a large performance margin on all three domains. Its audio encoder is jointly trained with visual and textual modalities through large-scale contrastive learning, producing representations that are both semantically informed and robust to content variability.
These results indicate that multimodally aligned audio features are crucial for accurately predicting human perceptual ratings.

\input{tables/ab_scale}
\subsubsection{Training scale}
Table~\ref{tab:ab_scale} examines how training set size affects overall PCC.
Performance improves steadily as the dataset grows from 50 to 100 hours, indicating that SAJ benefits from greater coverage of acoustic conditions and perceptual variations. Expanding to 150 hours further enhances music performance, which is an acoustically diverse and challenging domain, suggesting that additional data diversity is particularly valuable for non-speech categories.

Beyond this point, however, the results show saturation and mild fluctuations. For example, the 200-hour model underperforms the 150-hour counterpart on speech and music, reflecting that adding more data without careful balance may introduce distributional biases or increase label noise. After rebalancing and consolidating the dataset to 310 hours, performance reaches its best levels across all three modalities, matching or surpassing the previous peaks.

These findings suggest that scale alone is not sufficient; what matters is a sufficiently large and well-balanced dataset that captures the full range of perceptual phenomena relevant to audio separation. The strong performance at 310 hours confirms that our multi-stage data collection process yields a dataset that is both large enough and well-curated to support reliable perceptual modeling.

\input{tables/ab_joint}
\subsubsection{Separately vs. Jointly training}
Table~\ref{tab:ab_joint} compares separately trained models, where each metric (overall, recall, precision, faithfulness) is learned independently, with a jointly trained model that predicts all four dimensions simultaneously. The results show that joint training consistently improves performance across most modalities and metrics.

For the overall score, joint training yields clear gains for speech, music, and sound, indicating that leveraging shared perceptual cues across different dimensions helps the model form a more coherent representation of separation quality. Similar improvements are observed for precision, where joint training enhances performance across all three modalities, particularly for music and sound domains that have complex spectral structures.

While recall and faithfulness each show small modality-specific variations, the jointly trained model still matches or surpasses the separately trained version on most metrics. These fluctuations suggest minor trade-offs in isolated domains, but they are outweighed by the robust and consistent improvements on the global metrics that matter most for perceptual evaluation.

\input{tables/ab_proxy}
\subsubsection{Auxiliary Task}
Table~\ref{tab:ab_proxy} evaluates the impact of incorporating the auxiliary alignment detection task during training. Across all three modalities, the use of the auxiliary task leads to substantial improvements in overall PCC, showing relative gains of 2.3$\%$ (speech), 6.7$\%$ (music), and 6.3$\%$ (sound).

These consistent gains highlight the importance of teaching the model to first understand whether the separated audio actually follows the input prompt. By learning this binary alignment auxiliary task, the model develops stronger cross-modal grounding and becomes more sensitive to perceptually relevant mismatches between the prompt and the output audio. As a result, the SAJ model better captures the true perceptual quality reflected in human ratings.

\input{tables/ab_size}

\subsubsection{Model size}
Table \ref{tab:ab_size} compares different SAJ model sizes and the effect of using only the 4th Transformer block (L4).
Using only L4 consistently reduces performance, showing that full multi-layer encoder features are important for perceptual scoring.
SAJ-Light provides a strong lightweight alternative and outperforms its L4 variant, demonstrating good parameter efficiency.
However, SAJ-Base achieves the highest correlations across all modalities, indicating that the richer representations from the full PE-AV base encoder offer the strongest perceptual alignment.
Overall, SAJ-Light is efficient and competitive, but SAJ-Base remains the best-performing model when accuracy is the priority.

\subsection{Applications}

\begin{figure*}[t]
  \centering
  \includegraphics[width=0.9\linewidth]{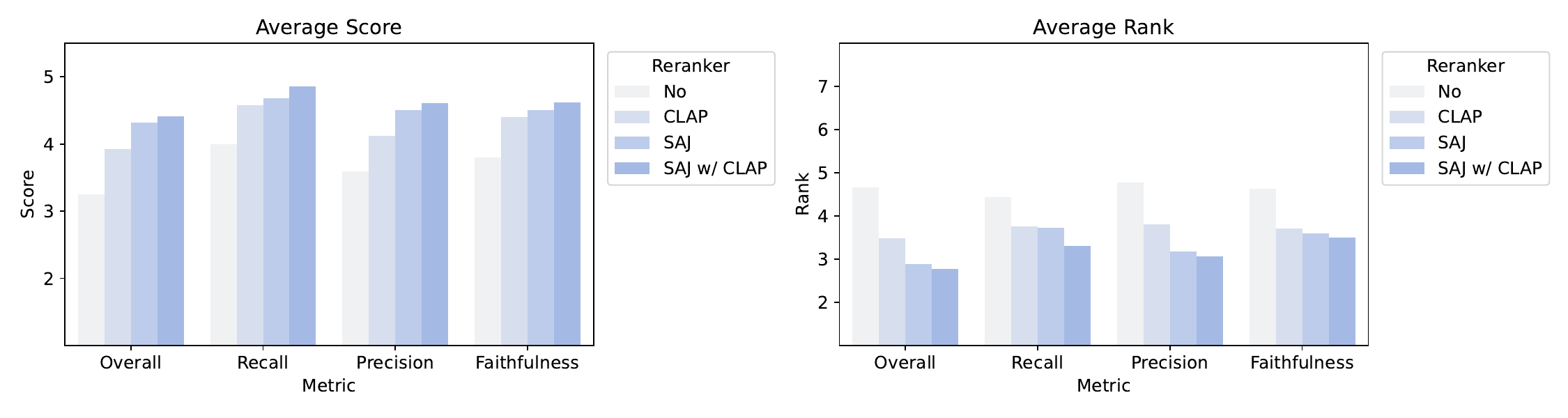}
  \caption{Separation performance comparison between different rerankers. We use 8 candidates for the reranking in SAM Audio~\cite{samaudio}.}
  \label{fig:saj_reranking}
\end{figure*}

\subsubsection{Reranking}
We evaluate the effectiveness of SAJ as a reranking module in audio separation by selecting the best output among eight candidate separation results per mixture. Experiments are conducted on the SAM Audio-Bench~\cite{samaudio} using the SAM Audio model. As shown in Fig.~\ref{fig:saj_reranking}, SAJ substantially improves the perceptual quality of the selected output across all four evaluation dimensions.

Compared with the “No Reranker’’ baseline, which chooses candidates arbitrarily, SAJ-ranked outputs achieve higher human-aligned scores, notably improving overall quality from roughly 3.2 to 4.2. SAJ also outperforms CLAP-based reranking, which struggles to distinguish fine-grained perceptual differences such as residual artifacts or missing target components.

In terms of ranking stability, SAJ yields the lowest average rank (best ranking) across all metrics, demonstrating its ability to consistently identify the best candidate among the eight separation outputs. Combining SAJ with CLAP brings only marginal improvement, indicating that SAJ already captures the majority of perceptually relevant cues.

These results highlight SAJ’s strong capability as a practical and reliable separation reranker, enabling significant perceptual gains without modifying the underlying separation models.

\subsubsection{Data Filtering and Pseudo Labeling}

\begin{figure}[t]
  \centering
  \includegraphics[width=0.7\linewidth]{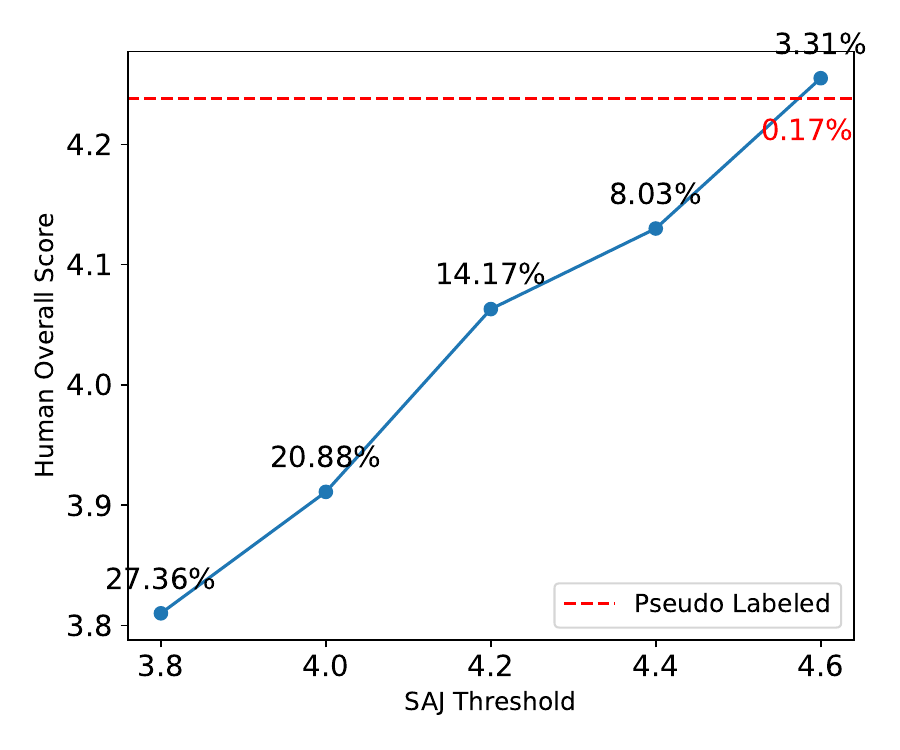}
  \caption{Human overall score as a function of the SAJ filtering threshold. Each point corresponds to retaining the top-$m\%$ highest-rated samples under SAJ, with the percentage indicating the remaining data after filtering. The red dashed line denotes the pseudo-labeled data used in SAM Audio~\cite{samaudio}.}
  \label{fig:saj_pl_raw}
\end{figure}

SAJ can be directly applied to data filtering and pseudo labeling in audio separation pipelines.

We randomly sample 10,000 mixture–separated pairs from the VGGSound and AudioSet-SL datasets. For five SAJ overall-score thresholds (3.8, 4.0, 4.2, 4.4, and 4.6), we filter the dataset accordingly and then randomly select 100 examples per threshold. Human judgment scores are collected following the procedure described in~\ref{sec:text_saj_data}.
For each threshold, we report the human overall score together with the percentage of data that remains after filtering.
We also compare against the pseudo-labeling strategy used in SAM Audio, which yields a human overall score of 4.26 while retaining only 0.17$\%$ of the data.

Fig~\ref{fig:saj_pl_raw} shows that SAJ scores are strongly aligned with human perceptual quality, making SAJ effective for both pseudo labeling and data filtering. As the SAJ threshold increases, the retained samples consistently achieve higher human overall scores, indicating that SAJ provides a reliable ranking of separation quality. Notably, for a similar target quality level, SAJ retains far more data than the pseudo-labeled set used in SAM Audio (red line), demonstrating its superior sample efficiency.

At the same time, the percentage markers illustrate that SAJ offers a controllable filtering mechanism: higher thresholds reduce dataset size but substantially improve perceptual quality. Thus, SAJ can be used to construct cleaner training subsets or to curate high-precision pseudo labels, supporting scalable and perceptually aligned data preparation for separation models.

\begin{figure}[t]
  \centering
  \includegraphics[width=0.7\linewidth]{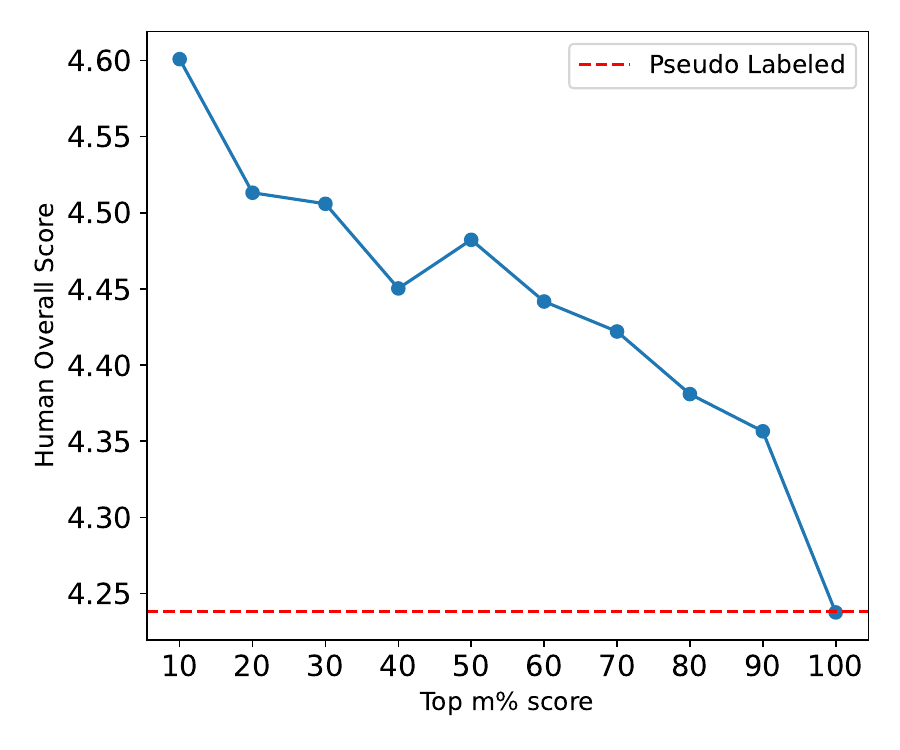}
  \caption{Quality of pseudo-labeled data after filtering with SAJ. For each top m$\%$ subset ranked by SAJ overall scores, we compute the corresponding human overall score. The red line marks the human score of the pseudo-labeled data used in SAM Audio~\cite{samaudio}.}
  \label{fig:saj_pl}
\end{figure}

Figure~\ref{fig:saj_pl} illustrates how SAJ can be used to refine the pseudo-labeled data employed in SAM Audio.
We randomly sample 500 examples from the pseudo-labeled dataset and obtain human evaluation scores following the procedure in Sec.~\ref{sec:text_saj_data}. We then rank the samples by their SAJ overall scores and compute the human overall score for the top m$\%$ subsets.

The results show a clear upward trend: as we retain smaller top-ranked portions according to SAJ, the corresponding human-rated quality increases consistently. Every filtered subset, regardless of the retention percentage, achieves higher human perceptual scores than the original pseudo-labeled data (red line).
This demonstrates that SAJ is highly effective at identifying high-quality pseudo-labels, making it a valuable tool for dataset refinement and for improving supervision quality in training separation models.

\subsubsection{Fine-grained Performance Analysis}
\input{tables/difficulty}
We can use SAJ to automatically sample evaluation cases with different levels of difficulty for any target audio concept list, thereby enabling a fully automatic and fine-grained performance analysis pipeline. To support this, we train a text-prompted SAJ model specifically to predict the intrinsic difficulty of a separation task. Since human annotations show that difficulty levels 4 and 5 are extremely sparse, we merge them and formulate a four-level difficulty scale (1–4). To stabilize training, the dataset is balanced with respect to the overall difficulty distribution.

Unlike the standard SAJ model, which predicts separation performance using both input and output audio, the difficulty model requires only the mixture audio and text prompt as input. This allows it to estimate how challenging a given separation instance is before running any separation model. 

We apply the predicted difficulty scores to the SAJ test set and compute the corresponding human performance ratings for the separated outputs produced by SAM Audio. Table \ref{tab:diff} shows how human-annotated separation performance varies with task difficulty. As expected, all performance metrics decrease as the difficulty level increases. Easy cases (Level 1) achieve the highest scores across all dimensions, while very difficult cases (Level 4) exhibit substantial degradation. By assigning difficulty scores to large pools of audio–text pairs, the system can automatically curate evaluation subsets at different difficulty levels (easy, moderate, hard, and very hard), fine-grained, concept-specific analysis of separation model robustness.

%% file: tables/saj_model.tex
\begin{table*}[t]
\centering
\caption{Model outputs used in the SAM Audio Judge DataSet.}
\adjustbox{max width=\textwidth}{%
\begin{tabular}{lcc}
\toprule
\textbf{Modality} & \textbf{Audio Separation Model}\\
\midrule
Speech & SAM Audio~\cite{samaudio}, MossFormer2~\cite{zhao2024mossformer2},  Tiger~\cite{xu2024tiger}, FastGeCo~\cite{wang2024noise}\\
Music& SAM Audio~\cite{samaudio}, AudioSep~\cite{AudioSep}, FlowSep~\cite{yuan2024flowsep}, ClapSep~\cite{CLAPSep}, SoloAudio~\cite{wang2025soloaudio}, Demucs~\cite{defossez2023htdemucs}, Spleeter~\cite{hennequin2020spleeter} \\
Sound& SAM Audio~\cite{samaudio}, AudioSep~\cite{AudioSep}, FlowSep~\cite{yuan2024flowsep}, ClapSep~\cite{CLAPSep}, SoloAudio~\cite{wang2025soloaudio}\\
\bottomrule
\end{tabular}
}
\label{tab:saj_model}
\end{table*}

%% file: tables/saj_data.tex
\begin{table}[t]
\centering
\caption{Audio samples and duration in SAJ DataSet}
\adjustbox{max width=.4\textwidth}{%
\begin{tabular}{lccc}
\toprule
\textbf{Split} & \textbf{Modality} & \textbf{Duration} & \textbf{Samples}\\
\midrule
\multirow{3}{*}{Training Set} & Speech &59.31 hrs&13,149\\
&Music &133.64 hrs&26,101\\
&Sound &117.52 hrs&37,444\\
\midrule
\multirow{3}{*}{Test Set} & Speech & 6.38 hrs &2,311\\
&Music &9.32 hrs&3,367\\
&Sound &31.72 hrs&11,476
\\
\bottomrule
\end{tabular}
}
\label{tab:saj_data}

\end{table}

%% file: tables/saj_com.tex
\begin{table*}[t]
\centering
\caption{Comparison Between SAM Audio Judge Model and Baselines}
\adjustbox{max width=\textwidth}{%
\begin{tabular}{l| cccc |cccc |cccc}
\toprule
 & \multicolumn{4}{c|}{\textbf{Speech}}
 & \multicolumn{4}{c|}{\textbf{Music}}
 & \multicolumn{4}{c}{\textbf{Sound}} \\
Model
& Overall & Recall & Precision & Faithfulness
& Overall & Recall & Precision & Faithfulness
& Overall & Recall & Precision & Faithfulness \\
\midrule
\rowcolor{gray!15}\multicolumn{13}{c}{Pearson Correlation Coefficient (PCC $\uparrow$)} \\
\midrule
CLAP &  0.490&  0.431& 0.283 &  0.477&  0.487&  0.416&  0.385&  0.432&  0.367&  0.431&  0.283&  0.418\\
AES-PC diff  & 0.227& -0.027&0.362&-0.060&0.111&-0.176&0.318&-0.155&0.182&-0.030&0.332&-0.045\\
SDR Estimator        & 0.336 & 0.004 & 0.403 & 0.055 &  0.369&  0.157&  0.388&  0.182&  0.181&  0.040&  0.222&  0.055\\
Gemini-2.5-pro         &  0.487& 0.498 &  0.169&  0.430 &0.351&0.287&0.115&0.303 &0.462&0.493&0.192&0.369\\
\textbf{SAM Audio Judge}        & \textbf{0.883} & \textbf{0.943} & \textbf{0.841} & \textbf{0.891} & \textbf{0.815} & \textbf{0.858} & \textbf{0.766} & \textbf{0.791} & \textbf{0.815} & \textbf{0.837} & \textbf{0.775} & \textbf{0.818} \\
\midrule
\rowcolor{gray!15}\multicolumn{13}{c}{Spearman Rank Correlation Coefficient (SRCC $\uparrow$)} \\
\midrule
CLAP &  0.380&  0.291& 0.325 &  0.273&  0.285&  0.293&  0.199&  0.296&  0.493&  0.376& 0.388 & 0.406 \\
AES-PC diff  & 0.231&-0.094& 0.360&-0.165&0.167&-0.108&0.323&-0.088&0.197&-0.076&0.373&-0.076\\
SDR Estimator        & 0.338 & 0.000 & 0.395 & 0.079 & 0.390 & 0.203 & 0.375 & 0.210 & 0.173 & 0.053 & 0.220 & 0.073 \\
Gemini-2.5-pro        &0.495&0.361&-0.015&0.117 &0.338&0.232&0.010&0.008 &0.390&0.324&-0.006&0.180
\\
\textbf{SAM Audio Judge}        & \textbf{0.817} & \textbf{0.573} & \textbf{0.774} & \textbf{0.573} & \textbf{0.714} & \textbf{0.569} & \textbf{0.658} & \textbf{0.476} & \textbf{0.781} & \textbf{0.660} & \textbf{0.734} & \textbf{0.607} \\
\bottomrule
\end{tabular}
}
\label{tab:res_saj_com}
\end{table*}

%% file: tables/ab_loss.tex
\begin{table}[h!]
\centering
\caption{Comparison between different training loss functions on overall PCC ($\uparrow$). No proxy task is applied.}
\adjustbox{max width=.3\textwidth}{%
\begin{tabular}{lccc}
\toprule
\textbf{Loss} & \textbf{Speech} & \textbf{Music} & \textbf{Sound}\\
\midrule
CE&0.589&0.544&0.528\\
KL-div&0.575&0.537&0.544\\
MAE&0.828&0.677&0.733\\
MSE&0.834&0.717&\textbf{0.772}\\
\textbf{MAE+MSE}&\textbf{0.863}&\textbf{0.741}&0.767\\
\bottomrule
\end{tabular}
}
\label{tab:ab_loss}

\end{table}

%% file: tables/ab_backbone.tex
\begin{table}[h!]
\centering
\caption{Comparison between different backbones on overall PCC ($\uparrow$). No proxy task is applied.}
\adjustbox{max width=.3\textwidth}{%
\begin{tabular}{lccc}
\toprule
\textbf{Backbone} & \textbf{Speech} & \textbf{Music} & \textbf{Sound}\\
\midrule
WavLM&0.640&0.528&0.533\\
DAC-VAE&0.681&0.565&0.600\\
\textbf{PE-AV}&\textbf{0.863}&\textbf{0.741}&\textbf{0.767}\\
\bottomrule
\end{tabular}
}
\label{tab:ab_backbone}

\end{table}

%% file: tables/ab_scale.tex
\begin{table}[h!]
\centering
\caption{Effect of training set size on overall PCC ($\uparrow$). Training data are collected incrementally using a three-stage pipeline (Batch 1--3).
All models are trained without proxy tasks.
All models are trained without proxy tasks. }
\adjustbox{max width=.4\textwidth}{%
\begin{tabular}{lccc}
\toprule
\textbf{Training Set} & \textbf{Speech} & \textbf{Music} & \textbf{Sound}\\
\midrule
50 hrs (Batch 1)&0.811&0.666&0.686\\
100 hrs&0.862&0.695&0.734\\
150 hrs&0.859&\textbf{0.745}&0.730\\
200 hrs (Batch 1+2)&0.847&0.725&0.765\\
250 hrs&\textbf{0.863}&0.732&0.760\\
\textbf{310 hrs (Batch 1+2+3)}&\textbf{0.863}&0.741&\textbf{0.767}\\
\bottomrule
\end{tabular}
}
\label{tab:ab_scale}

\end{table}

%% file: tables/ab_joint.tex
\begin{table}[h!]
\centering
\caption{Comparison between separately training and jointly training the four performance dimensions on PCC ($\uparrow$). No proxy task is applied.}
\adjustbox{max width=.35\textwidth}{%
\begin{tabular}{llccc}
\toprule
\textbf{Metric} & \textbf{Mode} & \textbf{Speech} & \textbf{Music} & \textbf{Sound}\\
\midrule
Overall&Separately&0.849&0.718&0.731\\
&Jointly&\textbf{0.863}&\textbf{0.741}&\textbf{0.767}\\
\midrule
Recall&Separately&0.879&0.683&\textbf{0.773}\\
&Jointly&\textbf{0.919}&\textbf{0.688}&0.764\\
\midrule
Precision&Separately&0.823&0.668&0.702\\
&Jointly&\textbf{0.846}&\textbf{0.702}&\textbf{0.729}\\
\midrule
Faithfulness&Separately&\textbf{0.872}&0.623&0.720\\
&Jointly&0.859&\textbf{0.667}&\textbf{0.741}\\
\bottomrule
\end{tabular}
}
\label{tab:ab_joint}

\end{table}

%% file: tables/ab_proxy.tex
\begin{table}[h!]
\centering
\caption{Comparison using and without using auxiliary task on overall PCC ($\uparrow$).}
\adjustbox{max width=.35\textwidth}{%
\begin{tabular}{lccc}
\toprule
\textbf{Mode} & \textbf{Speech} & \textbf{Music} & \textbf{Sound}\\
\midrule
w/o auxiliary task&0.863&0.741&0.767\\
w/ auxiliary task&\textbf{0.883}&\textbf{0.791}&\textbf{0.815}\\
\bottomrule
\end{tabular}
}
\label{tab:ab_proxy}

\end{table}

%% file: tables/ab_size.tex
\begin{table}[h!]
\centering
\caption{Comparison between different SAJ models on overall PCC ($\uparrow$).}
\adjustbox{max width=.4\textwidth}{%
\begin{tabular}{lcccc}
\toprule
\textbf{Model} & \textbf{\#Parameters}& \textbf{Speech} & \textbf{Music} & \textbf{Sound}\\
\midrule
SAJ-Light-L4&0.52B&0.863&0.787&0.793\\
SAJ-Light&0.57B &0.870&0.798&0.793\\
SAJ-Base-L4& 0.78B&0.843&0.779&0.748\\
SAJ-Base & 1.47B&\textbf{0.883}&\textbf{0.791}&\textbf{0.815}\\
\bottomrule
\end{tabular}
}
\label{tab:ab_size}

\end{table}

%% file: tables/difficulty.tex
\begin{table}[h!]
\centering
\caption{Human-rated separation quality across different task difficulty levels.}
\adjustbox{max width=.45\textwidth}{%
\begin{tabular}{ccccc}
\toprule
\textbf{Difficulty Level} & \textbf{Overall} & \textbf{Recall} & \textbf{Precision}& \textbf{Faithfulness}\\
\midrule
1&3.716&4.084& 3.984& 3.963\\
2&3.327&3.995&3.537&3.798\\
3&3.272&4.022&3.326&3.807\\
4&2.894&3.791&3.024&3.528\\
\bottomrule
\end{tabular}
}
\label{tab:diff}

\end{table}

%% file: multimodal_saj.tex
\section{Multimodal SAJ}
\input{tables/saj_data_joint}

\input{tables/joint_res}

\subsection{Data collection}
To train a multimodal SAJ model, we require data that covers multiple input prompt types.
Following~\cite{samaudio}, we construct a training set supporting text, visual, and span prompts, with the goal of transferring the text-prompted SAJ into a fully multimodal version.
Our data collection pipeline consists of four stages:
\begin{itemize}
    \item \textbf{Stage 1: Generate pseudo-labeled data}: Following~\cite{samaudio}, we apply pseudo-labeling to large-scale audio–visual in-house data to obtain input mixtures and corresponding clean stems. PLM-Audio~\cite{samaudio} is first used to produce text descriptions, which then prompt the SAM Audio model to generate target and residual audio tracks. We apply several quality filters: (1) compute text–target similarity using CLAP and discard samples below 0.35; (2) compute text–residual similarity using CLAP and discard samples above 0.0; (3) compute AES-PC on the target audio and discard samples above 2.5; (4) detect overly silent outputs using a VAD~\cite{pydub} and discard samples with a silence ratio above 0.95. This yields mixtures, clean stems, and text prompts. Next, we generate visual masks by prompting SAM3~\cite{carion2025sam3segmentconcepts} with the text prompts, and obtain span prompts by applying VAD to the clean stems. Finally, we retain samples whose visual mask coverage ratio is between 0.02 and 0.8 and whose ImageBind~\cite{girdhar2023imagebind} similarity (between target audio and masked region) exceeds 0.2.
    \item \textbf{Stage 2: Generate mixture–separated pairs}: Instead of relying solely on mixtures from Stage 1, we also simulate mixtures using the clean stems, following~\cite{samaudio}. This simulated data enables instance-level separation tasks—e.g., separating one of two visually similar speakers—which are difficult to prompt using real mixtures.
    We then run SAM Audio to produce mixture–separated pairs with text, visual, and span prompts. No reranking is applied, allowing us to maintain a relatively balanced distribution of separation performance.
    \item \textbf{Stage 3: Generate SAJ scores}: We apply the text-prompted SAJ model to obtain SAJ scores using the input mixtures, separated outputs, and text prompts.
    \item \textbf{Stage 4: Post-processing}: We remove low-quality or inconsistent samples using the following rules: (1) discard samples with clean-stem–to–output CLAP similarity above 0.6 and SAJ overall score below 3.0; (2) discard samples with CLAP similarity below 0.4 and SAJ overall score above 3.0. We then balance the dataset according to SAJ overall scores so that the five score levels (1–5) are evenly represented in the final training set.
\end{itemize}

We collect a total of 1,133 hours of multimodal training data, with detailed statistics shown in Table~\ref{tab:saj_data_joint}.
For real-world evaluation, we use the side-by-side absolute category ratings from SAM Audio-Bench~\cite{samaudio}, which cover a diverse range of separation models and audio sources.

\subsection{Model training}

Fig~\ref{fig:saj_model} illustrates the multimodal SAJ architecture. During training, we condition the model on seven different combinations of prompt inputs:
(1) text only;
(2) visual only;
(3) span only;
(4) text and visual;
(5) text and span;
(6) visual and span;
(7) text, visual and span.

Both the pretraining and fine-tuning stages are trained for 300K updates with a constant learning rate of $3\times10^{-4}$, preceded by a 5K-step linear warmup. Other setups are the same with the text-prompted SAJ.

\subsection{Comparison with other metrics}

Table~\ref{tab:res_joint} reports the performance of SAJ and baseline models across speech, music, and sound. Unimodal baselines (CALP, ImageBind and SpanIou) exhibit very weak correlations with human judgments, highlighting that single-modality similarity measures are insufficient for assessing perceptual separation quality. In contrast, SAJ yields substantial improvements even when conditioned on a single prompt type. Among these, text-prompted SAJ performs strongest overall, particularly for music, while span-prompted SAJ achieves the best results for sound, reflecting the complementary nature of different prompt modalities.

Training SAJ jointly on all prompt combinations further enhances performance. The text-conditioned version of multimodal SAJ achieves the highest correlations for speech and competitive results for music and sound, showing that multimodal training provides beneficial cross-modal grounding. Visual- and span-conditioned multimodal SAJ also outperform or match their single-prompt counterparts, indicating improved robustness across prompt types. SRCC follow the same trends, confirming consistent rank-order alignment with human ratings.

\subsection{Performance on different prompts}
\input{tables/joint_res_joint}

Table~\ref{tab:res_joint_com} examines the impact of different prompt configurations on multimodal SAJ. Text prompts provide the strongest single-modality signal across most metrics, while visual and span cues offer complementary benefits: visual prompts contributing spatial grounding and span prompts improving temporal alignment. Combining prompts consistently boosts robustness, with Text + Span and Text + Visual outperforming their single-prompt counterparts in many cases. The best overall results are achieved by the Text + Visual + Span configuration, which integrates semantic, spatial, and temporal information and delivers the most reliable correlations across speech, music, and sound. These results demonstrate that multimodal prompting enhances SAJ’s perceptual alignment beyond what any single prompt modality can provide.

%% file: tables/saj_data_joint.tex
\begin{table}[t]
\centering
\caption{Audio samples and duration in Mulimodal SAJ DataSet}
\adjustbox{max width=.4\textwidth}{%
\begin{tabular}{lccc}
\toprule
\textbf{Split} & \textbf{Modality} & \textbf{Duration} & \textbf{Samples}\\
\midrule
\multirow{3}{*}{Training Set} & Speech & 288.85 hrs &115,242\\
&Music &297.22 hrs&112,891\\
&Sound &546.85 hrs&196,867\\
\midrule
\multirow{3}{*}{Test Set} & Speech & 7.40 hrs &2,823\\
&Music & 7.09 hrs&2,590\\
&Sound & 8.85 hrs&3,225\\
\bottomrule
\end{tabular}
}

\label{tab:saj_data_joint}

\end{table}

%% file: tables/joint_res.tex
\begin{table*}[t]
\centering
\caption{Comparison between SAJ and baseline models.}
\label{tab:res_joint}
\adjustbox{max width=\textwidth}{%
\begin{tabular}{l|l| cccc |cccc |cccc}
\toprule
 \multirow{2}{*}{\textbf{Method}}& \multirow{2}{*}{\textbf{Prompt}} & \multicolumn{4}{c|}{\textbf{Speech}}
 & \multicolumn{4}{c|}{\textbf{Music}}
 & \multicolumn{4}{c}{\textbf{Sound}} \\
\cmidrule(lr){3-6}\cmidrule(lr){7-10}\cmidrule(lr){11-14}
& 
& Overall & Recall & Precision & Faithfulness
& Overall & Recall & Precision & Faithfulness
& Overall & Recall & Precision & Faithfulness \\
\midrule
\rowcolor{gray!15}\multicolumn{14}{c}{Pearson Correlation Coefficient (PCC $\uparrow$)} \\
\midrule
CLAP~\cite{wu2023large} & Text & 0.274 &- &- &- &0.347 &- &- &- &0.351 &- &- &- \\
ImageBind~\cite{girdhar2023imagebind} & Visual & 0.095 &- &- &- &0.119 &- &- &- &0.109 &- &- &- \\
SpanIoU~\cite{samaudio} & Span & 0.363 &- &- &- &0.228 &- &- &- &0.187 &- &- &-\\
\midrule
\textbf{Single prompted SAJ}& Text& 0.680 & 0.601 & 0.737 & 0.397 & \textbf{0.721} & 0.856 & \textbf{0.758} & \textbf{0.756} & 0.536 & \textbf{0.617} & 0.664 & \textbf{0.532} \\
&Visual& 0.568 & 0.381 & 0.696 & 0.282 & 0.691 & 0.797 & 0.722 & 0.701 & 0.439 & 0.422 & 0.622 & 0.344\\
&Span& 0.663 & 0.578 & 0.753 & 0.462 & 0.496 & 0.480 & 0.525 & 0.427 & \textbf{0.560} & 0.583 & \textbf{0.670} & 0.502\\
\midrule
\textbf{Multiple prompted SAJ}& Text& \textbf{0.724} & \textbf{0.616} & \textbf{0.778} & \textbf{0.507} & 0.716 & \textbf{0.863} & 0.743 & 0.747 & 0.544 & 0.588 & 0.663 & 0.488 \\
&Visual& 0.606 & 0.413 & 0.690 & 0.321 & 0.688 & 0.798 & 0.707 & 0.698 & 0.440 & 0.437 & 0.596 & 0.328\\
&Span& 0.646 & 0.497 & 0.727 & 0.424 & 0.555 & 0.531 & 0.594 & 0.507 & 0.503 & 0.543 & 0.599 & 0.446\\
\midrule
\rowcolor{gray!15}\multicolumn{14}{c}{Spearman Rank Correlation Coefficient (SRCC $\uparrow$)} \\
\midrule
CLAP~\cite{wu2023large} & Text &0.259 &- &- &- &0.322 &- &- &- &0.36 &- &- &-\\
ImageBind~\cite{girdhar2023imagebind} & Visual & 0.097 &- &- &- &0.097 &- &- &- &0.107 &- &- &- \\
SpanIoU~\cite{samaudio} & Span & 0.366 &- &- &- &-0.064 &- &- &- &0.193 &- &- &-\\
\midrule
\textbf{Single prompted SAJ}& Text& 0.690 & 0.437 & 0.743 & 0.297 & \textbf{0.714} & 0.611 & \textbf{0.750} & 0.528 & 0.528 & 0.496 & \textbf{0.668} & \textbf{0.456} \\
&Visual& 0.599 & 0.400 & 0.704 & 0.291 & 0.685 & 0.590 & 0.721 & 0.509 & 0.441 & 0.407 & 0.614 & 0.345\\
&Span& 0.666 & \textbf{0.457} & 0.747 & 0.323 & 0.485 & 0.464 & 0.519 & 0.328 & \textbf{0.558} & \textbf{0.506} & 0.666 & 0.428\\
\midrule
\textbf{Multiple prompted SAJ}& Text& \textbf{0.737} & 0.455 & \textbf{0.792} & \textbf{0.371} & 0.705 & \textbf{0.616} & 0.733 & \textbf{0.532} & 0.546 & 0.485 & 0.657 & 0.422 \\
&Visual& 0.624 & 0.370 & 0.692 & 0.298 & 0.683 & 0.604 & 0.705 & 0.505 & 0.424 & 0.458 & 0.560 & 0.359\\
&Span& 0.651 & 0.415 & 0.731 & 0.346 & 0.526 & 0.462 & 0.583 & 0.418 & 0.506 & 0.474 & 0.591 & 0.414\\
\bottomrule
\end{tabular}
}
\end{table*}

%% file: tables/joint_res_joint.tex
\begin{table*}[t]
\centering
\caption{Comparison between different prompts in Multiple prompted SAJ.}
\label{tab:res_joint_com}
\adjustbox{max width=\textwidth}{%
\begin{tabular}{l| cccc |cccc |cccc}
\toprule
 \multirow{2}{*}{\textbf{Prompt}} & \multicolumn{4}{c|}{\textbf{Speech}}
 & \multicolumn{4}{c|}{\textbf{Music}}
 & \multicolumn{4}{c}{\textbf{Sound}} \\
\cmidrule(lr){2-5}\cmidrule(lr){6-9}\cmidrule(lr){10-13}
& Overall & Recall & Precision & Faithfulness
& Overall & Recall & Precision & Faithfulness
& Overall & Recall & Precision & Faithfulness \\
\midrule
\rowcolor{gray!15}\multicolumn{13}{c}{Pearson Correlation Coefficient (PCC $\uparrow$)} \\
\midrule
Text& 0.724 & 0.616 & 0.778 & 0.507 & \textbf{0.716} & \textbf{0.863} & \textbf{0.743} & \textbf{0.747} & 0.544 & \textbf{0.588} & 0.663 & \textbf{0.488} \\
Visual& 0.606 & 0.413 & 0.690 & 0.321 & 0.688 & 0.798 & 0.707 & 0.698 & 0.440 & 0.437 & 0.596 & 0.328\\
Span& 0.646 & 0.497 & 0.727 & 0.424 & 0.555 & 0.531 & 0.594 & 0.507 & 0.503 & 0.543 & 0.599 & 0.446\\
Text + Visual & 0.726 & 0.609 & 0.778 & 0.495 & \textbf{0.716} & 0.860 & \textbf{0.743} & 0.745 & 0.545 & 0.578 & 0.668 & 0.474\\
Text + Span & 0.730 & \textbf{0.626} & 0.782 & \textbf{0.512} & 0.704 & 0.840 & 0.734 & 0.729 & 0.547 & 0.588 & 0.663 & 0.486\\
Visual + Span & 0.663 & 0.513 & 0.725 & 0.412 & 0.677 & 0.783 & 0.704 & 0.686 & 0.454 & 0.427 & 0.601 & 0.346\\
Text + Visual + Span & \textbf{0.733} & 0.622 & \textbf{0.783} & 0.506 & \textbf{0.716} & 0.858 & \textbf{0.743} & 0.746 & \textbf{0.555} & 0.583 & \textbf{0.667} & 0.477\\
\midrule
\rowcolor{gray!15}\multicolumn{13}{c}{Spearman Rank Correlation Coefficient (SRCC $\uparrow$)} \\
\midrule
Text& 0.737 & 0.455 & 0.792 & 0.371 & 0.705 & \textbf{0.616} & 0.733 & 0.532 & 0.546 & 0.485 & 0.657 & \textbf{0.422} \\
Visual& 0.624 & 0.370 & 0.692 & 0.298 & 0.683 & 0.604 & 0.705 & 0.505 & 0.424 & 0.458 & 0.560 & 0.359\\
Span& 0.651 & 0.415 & 0.731 & 0.346 & 0.526 & 0.462 & 0.583 & 0.418 & 0.506 & 0.474 & 0.591 & 0.414\\
Text + Visual & 0.677 & 0.426 & 0.732 & 0.372 & 0.670 & 0.588 & 0.703 & 0.498 & 0.454 & 0.463 & 0.582 & 0.373\\
Text + Span & 0.744 & 0.458 & 0.795 & 0.378 & 0.698 & 0.609 & 0.726 & 0.517 & 0.546 & 0.487 & 0.656 & 0.417\\
Visual + Span & 0.742 & 0.464 & 0.789 & 0.380 & 0.705 & 0.606 & 0.735 & 0.527 & 0.539 & 0.493 & \textbf{0.662} & 0.413\\
Text + Visual + Span & \textbf{0.746} & \textbf{0.466} & \textbf{0.793} & \textbf{0.385} & \textbf{0.707} & 0.610 & \textbf{0.736} & \textbf{0.534} & \textbf{0.550} & \textbf{0.500} & 0.660 & 0.415\\
\bottomrule
\end{tabular}
}
\end{table*}

%% file: conclusion.tex
\section{Conclusions}

This work introduces SAJ, a perceptually aligned, reference-free, prompt-aware evaluation framework for open-domain audio separation. By leveraging large-scale human judgments, cross-modal pretraining, and unified text–visual–span prompting, SAJ moves beyond the limitations of distortion-based and unimodal similarity metrics, providing reliable assessments that closely track human perceptual preferences. Our experiments demonstrate that SAJ not only correlates strongly with human ratings across speech, music, and environmental sounds, but also enables practical applications such as high-quality data filtering, pseudo labeling, and separation reranking at scale. The multimodal SAJ further shows that integrating semantic, spatial, and temporal cues yields more comprehensive and robust evaluations than any single prompt modality.

Looking forward, SAJ opens several promising directions. As separation models continue to evolve toward richer multimodal conditioning and more complex real-world mixtures, SAJ offers a foundation for developing evaluation tools that are equally flexible and perceptually grounded. Extending SAJ toward interactive evaluation, user intent modeling, and cross-domain generalization may further close the gap between automatic metrics and human auditory perception. Ultimately, we hope SAJ contributes to a shift in audio research toward metrics that reflect not only signal fidelity, but the nuanced perceptual qualities that matter in real-world listening.

%% file: ack.tex
\section{Acknowledgements}
The authors would like to thank Matt Le, Julius Richter, Sanyuan Chen, Heng-Jui Chang, Luya Gao, Dangna Li, Cynthia Gao and Carleigh Wood.